\documentclass[fleqn,usenatbib]{mnras}

\usepackage{graphicx}	
\usepackage{amsmath}	
\usepackage{amssymb}	
\usepackage{multicol}        
\usepackage{bm}		
\usepackage{pdflscape}	

\usepackage[T1]{fontenc}
\usepackage{ae,aecompl}

\usepackage{newtxtext,newtxmath}
\def\SG{\textcolor{black}}
\def\CB{\textcolor{black}}

\title[Identifying galaxies with extreme levels of Lyman continuum leakage]{Identifying reionization-epoch galaxies with extreme levels of Lyman continuum leakage in James Webb Space Telescope surveys}

\author[S. K. Giri et al.]{Sambit K. Giri$^{1}$\thanks{Contact e-mail: \href{mailto:sambit.giri@astro.su.se}{sambit.giri@astro.su.se}}, Erik Zackrisson$^{2}$, Christian Binggeli$^{2}$, Kristiaan Pelckmans$^{3}$ and
\newauthor
Rub\'en 
Cubo$^{3}$ 
\\
$^{1}$Department of Astronomy and Oskar Klein Center, Stockholm University, AlbaNova, SE-106 91 Stockholm, Sweden \\
$^{2}$Observational Astrophysics, Department of Physics and Astronomy, Uppsala University, Box 516, SE-751 20 Uppsala, Sweden \\
$^{3}$ Department of Information Technology, Division of Systems and Control (Syscon), Uppsala University, Box 337, SE-751 05 Uppsala, Sweden
}

\date{Accepted 2019 December 03. Received 2019 December 03; in original form 2019 February 25}

\pubyear{2019}

\begin{document}
\label{firstpage}
\pagerange{\pageref{firstpage}--\pageref{lastpage}}
\maketitle

\begin{abstract}
The James Webb Space Telescope (JWST) NIRSpec instrument will allow rest-frame ultraviolet/optical spectroscopy of galaxies in the epoch of reionization (EoR). Some galaxies may exhibit significant leakage of hydrogen-ionizing photons into the intergalactic medium, resulting in faint nebular emission lines. We present a machine learning framework for identifying cases of very high hydrogen-ionizing photon escape from galaxies based on the data quality expected from potential NIRSpec observations of EoR galaxies in lensed fields. We train our algorithm on mock samples of JWST/NIRSpec data for galaxies at redshifts $z=6$--10. To make the samples more realistic, we combine synthetic galaxy spectra based on cosmological galaxy simulations with observational noise relevant for $z\gtrsim 6$ objects of a brightness similar to EoR galaxy candidates uncovered in Frontier Fields observations of galaxy cluster Abell-2744 and MACS-J0416. We find that ionizing escape fractions ($f_\mathrm{esc}$) of galaxies brighter than $m_\mathrm{AB,1500} \approx 27$ mag may be retrieved with mean absolute error $\Delta f_\mathrm{esc}\approx$0.09(0.12) for 24h (1.5h) JWST/NIRSpec exposures at resolution R=100. For 24h exposure time, even fainter galaxies ($m_\mathrm{AB,1500} < 28.5$ mag) can be processed with $\Delta f_\mathrm{esc}\approx$0.14. This framework simultaneously estimates the redshift of these galaxies with a relative error less than 0.03 for both 24h ($m_\mathrm{AB,1500} < 28.5$ mag) and 1.5h ($m_\mathrm{AB,1500} < 27$ mag) exposure times. We also consider scenarios where just a minor fraction of galaxies attain high $f_\mathrm{esc}$ and present the conditions required for detecting a subpopulation of high $f_\mathrm{esc}$ galaxies within the dataset.
\end{abstract}




\section{Introduction}
\label{intro}
In order for the galaxy population at redshift $z\gtrsim 6$ to explain the reionization of the Universe, some fraction of the hydrogen-ionzing radiation (Lyman continuum, hereafter LyC) produced by hot stars must be able to evade absorption by dust and neutral gas within these galaxies and escape into the intergalactic medium (IGM); \SG{see \citet{dayal2018earlygalaxy} for a recent review}. The LyC escape fraction $f_\mathrm{esc}$ required for cosmic reionization is currently estimated at $f_\mathrm{esc}\approx 0.01$--0.2 \citep[e.g.][]{Hartley16,Sun16,Bouwens16} if $f_\mathrm{esc}$ is assumed to remain constant for all galaxies throughout the reionization epoch. However, some simulations have predicted significant variation in the LyC escape properties across the $z\gtrsim 6$ galaxy population -- either due to mass-dependent $f_\mathrm{esc}$, redshift-dependent $f_\mathrm{esc}$ or
large temporal variations of $f_\mathrm{esc}$ for each galaxy  \citep[e.g.][]{Yajima14,Paardekooper15,Cen15,Xu16,Trebitsch17,Sharma17,Kimm17}. 

At $z\lesssim 4$, LyC photons escaping from galaxies can be detected directly and numerous individual cases have already been identified at $z<1$ \citep{Bergvall06,Leitet13,Borthakur14,Izotov16a,Izotov16b},
mostly with  $f_\mathrm{esc}\sim 0.01$-0.1, but also a couple with more extreme levels of leakage  \citep[$f_\mathrm{esc}\gtrsim 0.5$][]{Izotov18a,Izotov18b}. 
Claims of individual detections of extreme LyC leakage have also been made at $z\approx 2$--3 \citep{Vanzella16,Shapley16,Bian17,Vanzella18,Fletcher18}. However, it should be noted that the correction for IGM attenuation required to convert a LyC detection into an $f_\mathrm{esc}$ estimate at these higher redshifts varies by a factor of a few due to sightline-to-sightline variations in IGM opacity \citep[e.g.][]{Vasei16}, which makes it difficult to accurately pin down $f_\mathrm{esc}$ from any individual object. Another complication that afflicts all direct LyC observations throughout the $z=0$--4 range is that these only probe the LyC leakage that takes place in the direction of the observer. If the leakage is highly anisotropic \citep[as seen in the simulations of e.g.][]{Paardekooper15,Cen15}, then the $f_\mathrm{esc}$ inferred from individual objects may not translate well into the global $f_\mathrm{esc}$ that matters for reionization.

At even higher redshift, the opacity for LyC photons grows due to the increasingly neutral IGM, which reduces the observed LyC flux by a factor of $\gtrsim 10$ at $z\gtrsim 4$ \citep{Inoue14} and effectively prohibits direct studies of LyC radiation at these redshifts. 

So, how does one study LyC leakage for individual objects in the actual reionization era? There are currently several indirect methods with the potential to accomplish this. The strengths of absorption lines due to gas in the interstellar or circumgalactic medium can measure the covering fraction of gas along the line of sight and therefore the likely $f_\mathrm{esc}$ \citep[e.g][]{Jones13,Leethochawalit16}. However, methods of this type requires high signal-to-noise ratios (SNR) in the continuum measurements and can in the foreseeable future only be applied to the very brightest reionization-epoch galaxies. Also, anisotropic leakage may cause the line-of-sight $f_\mathrm{esc}$ derived this way to be significantly different from the global value. 

Diagnostics based on emission lines show more promise for fainter galaxies (and hence larger samples of objects). High values of optical [OIII]/[OII] emission line ratios may indicate density-bounded conditions in the interstellar medium and has successfully been used to find LyC leakers at low redshift \citep{Izotov16a,Izotov16b,Izotov18a,Izotov18b}. However, such signatures may not be present if LyC preferentially takes place through low-density channels (a.k.a. the picket-fence model) through the gas, and all objects with high [OIII]/[OII] galaxies do not seem to display detectable levels of leakage \citep{Naidu18}. The profile of the Ly$\alpha$ emission line has also emerged as an interesting diagnostic of LyC leakage \citep[e.g.][]{Verhamme17}, although the absorption of Ly$\alpha$ photons by the neutral IGM may severely complicate its use in the EoR. \citet{MasRibas17} has proposed a method to estimate $f_\mathrm{esc}$ using the radial surface brightness profile of Ly$\alpha$ or H$\alpha$ emission from $z\approx 6-7$ galaxies which may provide a better handle on the global $f_\mathrm{esc}$, but can once again only be applied to the brightest galaxies.

\citet{Zackrisson13,Zackrisson17} presented an alternative method which may be able to single out individual galaxies with extreme levels of LyC leakage ($f_\mathrm{esc}\gtrsim 0.5$) based on the integrated luminosity of emission lines compared to the continuum flux. Due to the relatively simple astrophysics underlying the strengths of hydrogen Balmer lines, these studies focused on H$\beta$ -- which can readily be detected up to $z\approx 9$ through JWST/NIRSpec spectroscopy. However, the spectral information from all detectable emission lines compared to the continuum can in principle be combined to improve the estimate provided that a good model of how their strengths vary with $f_\mathrm{esc}$ is available \citep{Jensen16}. This technique may also be able to provide an $f_\mathrm{esc}$ estimate that is closer to the global value than absorption-line methods, at least in the case where dust effects can be assumed to be minimal \citep[for a discussion on caveats, see][]{Zackrisson17}. Similar methods have recently been applied to the study of far-infrared emission lines detected with ALMA \citep{Inoue16,Tamura18} and photometrically measured rest-frame optical emission lines \citep{Castellano17}. 

Here, we continue the work of \citet{Jensen16} by applying machine-learning techniques to the study of JWST spectra for the purpose of identifying high-$f_\mathrm{esc}$ candidates among $z\approx 6$--10 galaxies based on the strength of emission lines compared to the rest-frame ultraviolet(UV)/optical continuum. While \citet{Jensen16} only studied the prospects to single out high-$f_\mathrm{esc}$ objects under idealized conditions (redshift $z=7$ and neglecting redshift uncertainties), we attempt to carry out a similar analysis for a mock sample covering a wide range in brightness and redshift. Our mock samples are designed to simulate JWST observations of the Frontier Fields\footnote{\url{https://frontierfields.org}} (hereafter called FF), thereby covering both strongly lensed and unlensed galaxies, and to use the setup selected for Guaranteed Time Observations of one of the FF in JWST cycle 1, as well as more ambitious observations that may be considered for later cycles.

In Section~\ref{samples}, we describe our procedure for creating mock galaxy samples with different distributions of LyC escape fractions and in Section~\ref{ML} we provide the details of the machine learning techniques we use to extract information from these. Our ability to accurately identify high-$f_\mathrm{esc}$ galaxies in the mock samples is analysed in Section~\ref{results}. Section~\ref{discussion} discusses the limitations of our current approach, along with procedures that could be improved. Our conclusions are summarised in Section~\ref{summary}.

\section{Mock samples}
\label{samples}
In this section, we \SG{describe our mock samples. We} discuss the FF catalogue used and the quantities that go into our simulation \SG{in Section~2.1}.  
\SG{In Section~2.2,} we explain the procedure for creating the sample of galaxy spectra meant to match the future observations of the JWST/NIRSpec instrument.

\subsection{ASTRODEEP catalogue}
The FF survey was an observational programme aimed at producing multiwavelength photometric catalogues of galaxies in the field of strong-lensing galaxy clusters. The survey accomplished this by simultaneously using the space-based (Hubble Space telescope and Spitzer Space Telescope) and ground-based (VLT) telescopes, thus obtaining a multi-filter photometric data set with wide wavelength coverage. The FF was performed as a pilot program to look for very faint galaxies that JWST would be able to study in more detail \citep[see][for details about the FF survey]{2017ApJ...837...97L}. The main targets were the six galaxy clusters Abell-2744, MACS-J0416, MACS-J0717, MACS-J1149, Abell-S1063 and Abell-370.

The FF data sets of the cluster and parallel fields of Abell-2744 (hereafter, A2744) and MACS-J0416 (hereafter, M0416) were analyzed to derive quantities such as the photometric redshifts and other rest frame properties of galaxy candidates in these fields as a part of the ASTRODEEP\footnote{\url{http://www.astrodeep.eu/}} programme \citep{2016A&A...590A..30M,2016A&A...590A..31C}. In this paper, we will use the publicly available ASTRODEEP catalogue\footnote{The catalogue was downloaded from \url{http://astrodeep.u-strasbg.fr/ff/index.html}} developed and described by \citet{2016A&A...590A..31C}. Table~\ref{tab:FF_catalogue} gives the number of galaxies observed in the each field along with their average magnification $\langle\mu \rangle$. 

\begin{table}
	\centering
	\caption{The number of photometrically selected $z\gtrsim 6$ galaxies in the ASTRODEEP Frontier Field catalogue and the number of simulated galaxies on which the algorithm is tested.}
	\label{tab:FF_catalogue}
	\begin{tabular}{lcccc} 
		\hline
		Field & $\langle\mu \rangle$ & ASTRODEEP & Testing set\\
		      &                      & galaxies  & galaxies\\
		\hline
		A2744-Cl & 2.77 & 127 & 2794 \\
		A2744-Par& 1.22 & 221 & 4862 \\
		M0416-Cl & 3.29 & 118 & 2596 \\
        M0416-Par& 1.15 & 178 & 3916 \\
		\hline
	\end{tabular}
\end{table}

\begin{figure}
  \centering
  \includegraphics[width=0.45\textwidth]{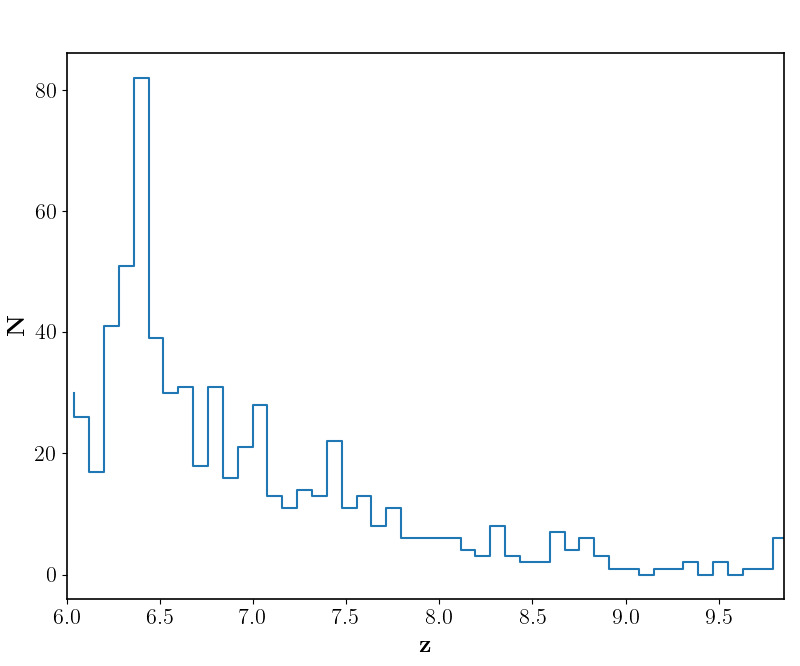}
  \caption{Histogram showing number of galaxies at each redshift in our sample selected from ASTRODEEP catalogue. The sample plotted here includes the galaxies from both fields. 
  }
  \label{fig:z_hist}
\end{figure}

From the ASTRODEEP catalogue, we select galaxies with $m_\mathrm{AB,1500}\approx 24-30$ and photometric redshifts $z\approx 6-10$. In Figure~\ref{fig:z_hist}, we plot the number of galaxies $N$ at various $z$ in our selected sample. We have considered galaxies from both cluster and parallel fields of the lensed galaxies in order to obtain a larger reference set when simulating the corresponding JWST spectra in order to the reduce the impact of cosmic variance on our procedure.

M0416 is one the most efficient lensing galaxy clusters \citep{2013ApJ...762L..30Z} and also among the targets of the JWST Cycle 1 Guaranteed Time Observation (GTO) {\it CAnadian NIRISS Unbiased Cluster Survey}\footnote{\url{http://www.stsci.edu/jwst/phase2-public/1208.pdf}}(CANCUS). A2477 has been selected as the target in the approved JWST Cycle 1 Early Release Science programme (ERS) proposal {\it Through the looking GLASS}\footnote{\url{http://www.stsci.edu/jwst/observing-programs/approved-ers-programs/program-1324}}. By choosing to simulate galaxies of similar brightness and redshift distribution as those fields, the procedure derived in the present paper can in principle be applied directly to the data obtained by these JWST programmes. In both these JWST surveys, the cluster fields will be targeted by NIRISS low-resolution grisms, NIRCam imaging and NIRSpec multi-object spectroscopy.

The exposure time ($t_\mathrm{exp}$) for the GTO proposal will be short ($\approx$ 1.5 h), whereas the ERS proposal will have a longer exposure time which will be more than 5 h (Tommaso Treu, private communication). 
Following \citet{Jensen16}, we consider a simplistic approach to simulate noise (see Section~\ref{sec:simulate_spectra}). Therefore we make a conservative choice and consider the exposure time $t_\mathrm{exp}=1.5$ h, which matches the GTO proposal, to roughly match the data quality expected from these programmes. Along with $t_\mathrm{exp}=1.5$ h, it is prudent to study a more optimistic scenario with $t_\mathrm{exp}=24$ h for future allocations. 
We should note that we have not considered the microshutter weights actually assigned to galaxies in the planned observations. We are simply considering the number of galaxies that could potentially be observed above a certain threshold brightness in these fields, not necessarily the ones that will be targeted by these programmes.

The FF will be observed with $R=100$ in the GTO programme but with $R=2700$ in the ERS case. In \citet{Jensen16}, the authors briefly discuss the effects of different spectral resolution settings, and find that the additional noise in the $R=1000$ spectra makes this mode less useful for the method discussed there. For the same reason, we choose to only consider $R=100$ spectra in this study. 

\subsection{Simulating the spectra}
\label{sec:simulate_spectra}

To generate synthetic spectra, we utilize galaxies from the CROC cosmological simulations developed by \cite{gnedin2014a} and \cite{gnedin2014b}. From this simulation, we extract star formation histories and metallicity distributions of 1300 galaxies with stellar masses $M_\mathrm{star}\geq 10^{7}M_{\odot}$ at redshifts $z\approx 7,8,9 \text{ and } 10$. We use synthetic spectra from the Yggdrasil spectral synthesis code \citep{zackrisson2011} in order to compute realistic spectra of the stellar population in each simulated galaxies. We use Yggdrasil with the BPASS v.2.0 models for binary stars \citep{eldridge2009} for metallicities $0.030 \geq Z\geq 0.001$ and the \cite{raiter2010} models for metal poor stars $Z = 10^{-5}$--$10^{-7}$. The nebular emission associated with each collection of stars is calculated using the Cloudy code \citep{Ferland13} under the assumption of an ionization-bounded nebula with holes free of gas and dust. We scale this nebular emission to account for different escape fractions ($f_\mathrm{{esc}}=0.0,0.10 \ldots 1.00$). 
For a more detailed description of the method or on the leakage geometry and its consequences, see \cite{Zackrisson13, Zackrisson17}. In order to account for dust, we use the dust recipe described in \cite{finlator2006} in combination with the \cite{pei1992} SMC dust reddening law for all galaxies. It should be noted that the mean UV slopes resulting from this procedure at $z\approx 7$ are characterised by \CB{$\beta \approx -2.5 \text{ to } -2.3$} which is slightly bluer than indicated by observations \citep[e.g.][]{bouwens14}. 

As explained in more detail in Section~\ref{ML}, the machine learning framework requires both a training set and a testing set of spectra. The training set contains the spectra whose properties are known. Therefore the training set is constructed from simulations. The framework will learn the relation between the spectra and the properties using the training set. Using this relation, we can extract the properties of the testing set, which serves as a proxy for actual JWST/NIRSpec observations. As actual JWST data are not yet available, we will create the testing set using the properties of our FF catalogue. Here, we explain how we produce our training and testing set.

\subsubsection{Training set}
\label{training}
For the training set we need spectra of galaxies over a wide range of redshifts ($z\approx 6-10$) and with high redshift resolution ($\Delta z=0.01$). We achieve this range and resolution by simply finding the closest match in redshift within the simulations. 
For the highest redshift set ($z\approx 10$), this corresponds to all galaxies that fulfil the stellar mass requirement adopted in \citet{Zackrisson17}. We have 22 galaxies in the highest redshift set. In order to avoid a bias towards any redshift, we randomly draw 22 galaxies from each set of lower redshifts. \CB{In order to make sure our training set consists of spectra with a similar noise level as what is expected for the FF galaxies, we scale the simulated spectra such that the magnitude range in the simulated galaxies roughly overlaps with the magnitudes observed in the FF.} 
For each of the 22 galaxies, we create synthetic spectra for cases with and without the Ly$\alpha$ emission line (see Section~\ref{unshifting}) and with varying $f_\mathrm{esc}$.

We then re-sample the spectra to match JWST/NIRSpec resolution and add simulated observational noise according to the recipe described in \cite{Jensen16}. This recipe utilises the NIRSpec spectral sensitivity curve, which gives us the minimum continuum flux observable with a SNR of 10, for a $10^4$ s exposure. The SNR in each spectral bin is calculated under the assumption that the wavelength-dependent flux sensitivty scales with the square root of the exposure time\footnote{\SG{It has been uncovered that \citet{Jensen16} overestimated SNR at the faint end.}}. The signal to noise in each spectral bin is then used to create random noise realisation while assuming that the noise in each spectral bin is Gaussian. Note that this ignores certain noise sources such as readout noise, and assumes that the galaxies can be treated as point sources. We perform this calculation for the two selected exposure times, 1.5 hours and 24 hours. This leaves us with a total of $\sim 180000$
simulated JWST/NIRSpec spectra of galaxies at redshifts $z\approx 6-10$ with magnitudes in the range $m_{1500,AB}\approx 24-30$ with observational noise and varying LyC escape fractions. 

\subsubsection{Testing set}



\CB{For the testing set, we do not follow the conventional method of keeping a part of the training set for testing purposes. Rather, we create the testing set independently by matching with the FF observations in order to understand how well our model works on realistic data. While the procedure is similar as for the training data (see section~\ref{training}), and the galaxies are drawn from the same simulation, in this case, we create a simulated spectrum for each FF galaxy. In order to do this, we bin the simulated galaxies at each redshift into magnitude bins. For each FF galaxy, a simulated spectrum is drawn from the bin that best matches the intrinsic (unlensed) magnitude of the observed galaxy. The simulated spectrum is then re-scaled and shifted to exactly match the observed magnitude and redshift of the object.}


We then re-sample the spectra to NIRSpec resolution and calculate the noise for each of the spectra following the procedure described above. This gives us simulated JWST/NIRSpec spectra with varying $f_\mathrm{esc}$ for each galaxy in our FF data. The number of simulated spectra for the testing set is given in Table~\ref{tab:FF_catalogue} from each field. \SG{We should point out that the number of simulated galaxies in the testing set is 22 times the number of galaxies in the ASTRODEEP catalogue because we consider 11 $f_\mathrm{esc}$ bins and we simulate spectra with and without the Ly$\alpha$ emission feature. One should not confuse this with the 22 galaxies found in the ASTRODEEP catalogue for the highest redshift set, which is described in the previous section.} While applying the machine learning procedure to real data, observations from the JWST/NIRSpec instrument will replace this testing set. The number of JWST/NIRSpec spectra simulated for each FF is 22 times the number of objects selected form the ASTRODEEP catalogue, based on the magnitude and redshift criteria used.

\section{Machine learning pipeline}
\label{ML}

\begin{figure}
  \centering
  \includegraphics[width=0.5\textwidth]{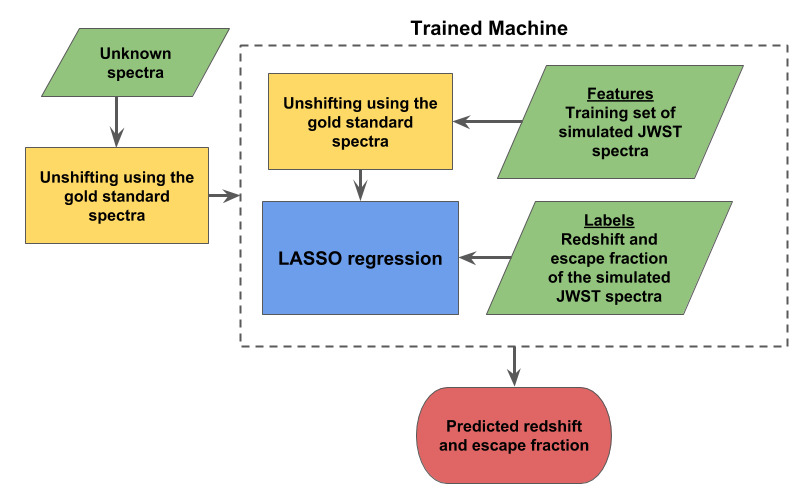}
  \caption{Flowchart of the machine learning pipeline. 
  The simulated (`mock') JWST spectra are \textit{unshifted} to the rest frame wavelength using GSS. 
  These spectra are used to train the model using LASSO. 
  When we get a new spectrum, we \textit{unshift} it and give it to the trained machine to predict its $z$ and $f_\mathrm{esc}$.}
  \label{fig:ML_flowchart}
\end{figure}

In this section, we describe our machine learning framework for predicting $z$ and $f_\mathrm{esc}$ of new JWST/NIRSpec spectra, which is illustrated in Figure~\ref{fig:ML_flowchart}. \SG{Before sending both the training and testing set spectra to our machine learning algorithm, we pre-process the spectra to bring them to the rest frame wavelength. We explain this pre-processing step in Section~\ref{unshifting}.}
Then the trained models are used to evaluate $z$ and $f_\mathrm{esc}$.
In Section~\ref{LASSO}, we describe the training algorithm used in this study.


\subsection{LASSO Regression}
\label{LASSO}


We use a machine learning algorithm known as \textit{least absolute shrinkage and selection operator} regression \citep[hereafter LASSO;][]{hastie01} to train our machine to predict $z$ and $f_\mathrm{esc}$. \citet{Jensen16} showed that LASSO can predict the $f_\mathrm{esc}$ from spectra of galaxies at known redshift. LASSO fits a linear model to the data set and controls the over-fitting by penalizing the goodness-of-fit term (training error). 
The prime advantage of LASSO is that it results in a sparse solution, indicating the few features in the candidate set that were actually relevant to the predictions. A sparse solution refers to a solution which contains often coefficients equal to zero.

In a machine learning framework, there are two data sets: the \textit{training} and \textit{testing} sets \citep{hastie01}. 
The model is trained on the training set, and then this model is used to evaluate the testing set. In the present case, we consider the simulated spectra as the training set and apply the model to mock observations. 
Let there be $m$ training samples: ($\mathbf{x}_1, y_1$), ($\mathbf{x}_2, y_2$), ... ($\mathbf{x}_m, y_m$). 
The input vector $\mathbf{x}_r$ contains $N$ features. $y_r$'s are the corresponding output values. 
A linear fit to the data set has the following form,
\begin{equation}
\hat{y} = \beta_0 + \sum_{i=1}^{N}\beta_ix_i
\end{equation}
where the $\beta$'s are the fitted coefficients (or parameters). The Gauss-Markov theorem states that minimizing the least square error is an optimal choice to determine the $\beta$'s \citep{hastie01} under certain assumptions. However, straightforward application of least squares when $m$ is large, can lead to over-fitting of the data \citep{hastie01}. In order to prevent this, a regularization term is added to the least square error. The modified estimator is given as
\begin{equation}
\sum_{j=1}^{m}[\hat{y}_j-y_j]^2 + \lambda\sum_{i=1}^{N}|\beta_i|
\end{equation}
where the second term is known as the regularisation term and $\lambda>0$ controls the trade-off. Minimising this estimator trains the model. Cross-validation can be used to determine this trade-off term $\lambda>0$ \citep{kohavi1995study, hastie01}. We follow the scheme described in \citet{Jensen16} with 10-fold cross-validation. For this study, we have used the tools provided in the {\sc scikit-learn} python package \citep{pedregosa2011scikit}. 

\subsection{Unshifting}
\label{unshifting}
As our data set contains spectra from different $z$, each nebular emission line will be observed at a different wavelength. The galaxies with low $f_\mathrm{esc}$  will typically have more prominent lines \citep{Zackrisson17} and \citet{Jensen16} showed that the prediction of $f_\mathrm{esc}$ depends strongly on the fluxes of these emission lines. For any machine learning algorithm to understand this dependence, each spectral feature must lie in the same wavelength bin. Therefore, we shift all the spectra to the rest frame wavelength before feeding those to our machine learning code. Throughout this paper, we will refer to this process as unshifting.

For unshifting, we use Gold Standard Spectra (hereafter, GSS), which are noise-free spectra at the best possible wavelength resolution in the rest frame. The wavelength sampling of JWST/NIRSpec instrument is not uniform. If we shift the spectra in the JWST/NIRSpec wavelength sampling, then the shift for each spectral feature will have a different value in wavelength units even though it is the same in bin units. 
Therefore, we use linear interpolation to obtain a uniform wavelength resolution with the bin-width roughly similar to that of the JWST/NIRSpec. The observed spectra are shifted to the rest frame by moving them until the cross-correlation with GSS is maximised. We record the shift for each observed spectrum and pass it on to our machine learning algorithm for $z$ and $f_\mathrm{esc}$ estimation.

The spectra from galaxies before the completion of reionization will be strongly affected by the presence of neutral hydrogen in the IGM. The fluxes at wavelengths less than that of Ly$\alpha$ line will be absorbed due to the Gunn-Peterson effect \citep{gunn1965density}. Even the strength of the Ly$\alpha$ emission line will be affected by the absorption in the damping wings of Gunn-Peterson absorption \citep{1998ApJ...501...15M}. Therefore, we consider a dictionary containing GSS with a full, 75 per cent, 50 per cent, 25 per cent and no Ly$\alpha$ emission line. This dictionary also contains GSS with various $f_\mathrm{esc}$. The shifting process is repeated for all the GSS in the dictionary because the result will be more accurate if the observed spectrum shifts to a GSS that has similar features. For example, an observed spectrum with no Ly$\alpha$ emission line and prominent nebular emission lines will provide the largest cross-correlation coefficient with a GSS where Ly$\alpha$ is absent and which has a low $f_\mathrm{esc}$.

Figure~\ref{fig:ML_flowchart} shows the flowchart of the entire machine learning pipeline. First, the spectra for training the machine are unshifted. The fluxes of the rest-frame spectra along with the shift values are given as input ($\mathbf{x}_r$) to LASSO. The framework trains two  models to the training data set: one models $z$ and the other one models $f_\mathrm{esc}$. The spectra from the testing set have to be unshifted before the models trained by LASSO can be used for prediction.

\section{Results}
\label{results}
In this section, we illustrate the performance of our procedure on the testing set, which contains mock spectra closely matched to the expected JWST/NIRSpec data quality for galaxies in FF catalogue of the lensed fields of A2744 and M0416. The performance of the method is better when the training and testing set have the same noise level \citep{Jensen16}. The SNR will vary with the redshift of the galaxy. However the SNR of the spectra will be similar for a particular exposure time \citep[see figure 2 in][]{Jensen16}. Therefore we consider the data set with the same exposure time to train and test the method.

\subsection{Redshift estimation}
\label{sec:red_est}

\begin{figure*}
  \centering
  \includegraphics[width=0.9\textwidth]{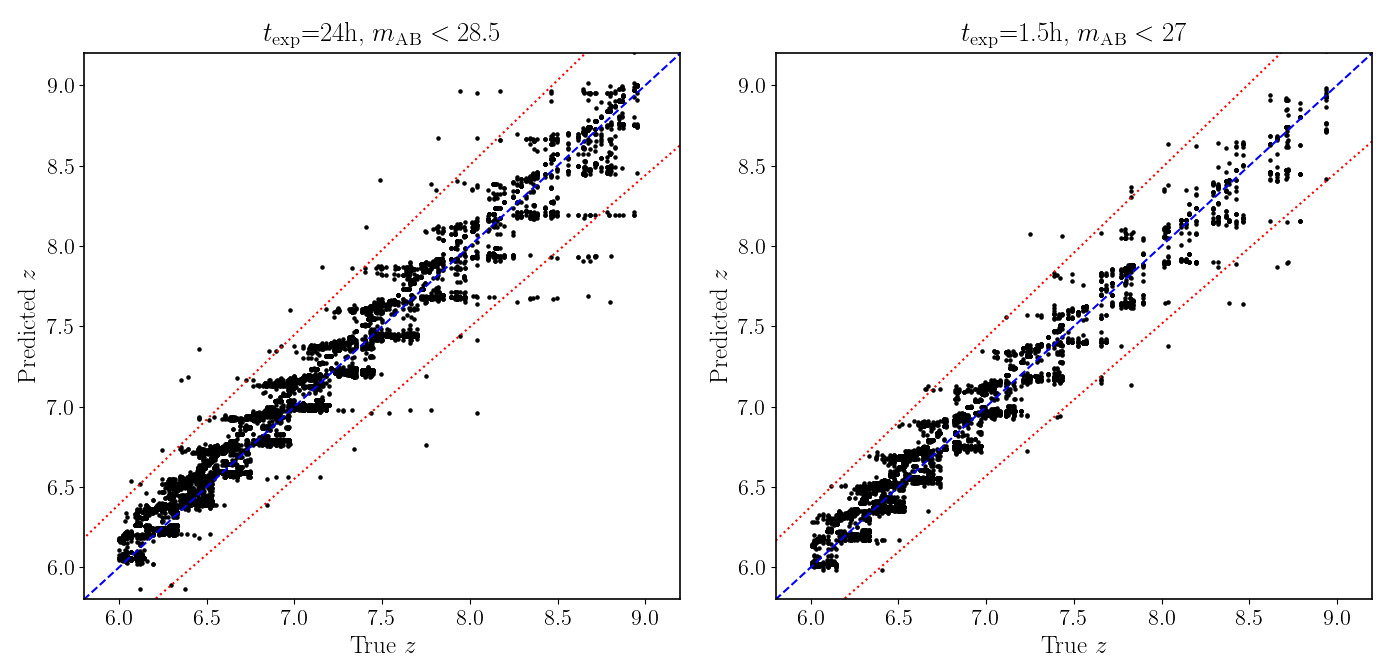}
  \caption{Redshifts predicted ($z^\mathrm{pred}$) by the machine learning pipeline for both the exposure times ($t_\mathrm{exp}$) considered in this work. The blue dashed line represents $z^\mathrm{true}$ = $z^\mathrm{pred}$. The red dashed lines enclose the region satisfying $\left | \frac{\Delta z}{1+z} \right |<0.06$. The points outside this region are considered outliers.
  We have selected spectra with SNR$\gtrsim 5$ for both data sets. 
  The relative error of the $z$ prediction is similar when the data set contains similar noise level. 
  Please note that due to significant overlap of data points close to the $z^\mathrm{true}$ = $z^\mathrm{pred}$ relation, this plot make the extreme outliers seem more likely than they really are.}
  \label{fig:z_est}
\end{figure*}

After training, LASSO has high values for the coefficients associated with the shift value and low values for those associated with the fluxes, which means, not unexpectedly, that the value $z$ strongly depends on the shift value. Therefore, more accurate unshifting will lead to better $z$ prediction. In order to estimate the scatter of the predicted $z$ against the true $z$, we use
the relative error 
\begin{equation}
\frac{\Delta z}{(1+z)} =  \frac{z^\mathrm{true}-z^\mathrm{pred}}{1+z^\mathrm{true}} \,
\end{equation}
where $z^\mathrm{true}$ and $z^\mathrm{pred}$ are the true and predicted redshifts. This quantity provides a good diagnostic for the accuracy of the redshift estimate  \citep[e.g.][]{dahlen2013critical,2016A&A...590A..31C}.

\begin{table}
	\centering
	\caption{The summary of $z$ of the testing set spectra predicted using our machine learning pipeline.}
	\label{tab:z_pred_summary}
	\begin{tabular}{lcccc} 
		\hline
		Testing Set & $\langle \frac{\Delta z}{1+z} \rangle$ & $\sigma_{\frac{\Delta z}{1+z}}$ & \% of outliers \\
		\hline
		$t_\mathrm{exp}=24$h, \SG{$m_\mathrm{AB,1500}<28.5$} & \SG{-0.005} &  \SG{0.018} & \SG{0.98} \\
		$t_\mathrm{exp}=1.5$h, \SG{$m_\mathrm{AB,1500}<27.0$} & \SG{-0.002} &  \SG{0.018} & \SG{0.86} \\
		\hline
	\end{tabular}
\end{table}

In Figure~\ref{fig:z_est}, we show the prediction of $z$ from the spectra with both $t_\mathrm{exp}=1.5$ and 24 h using a scatter plot of true redshift ($z^\mathrm{true}$) versus predicted redshift ($z^\mathrm{pred}$). Here, the optimal outcome ($z^\mathrm{true}$ = $z^\mathrm{pred}$) is indicated by a blue-dashed line. 
The spectra from the faintest galaxies are not unshifted properly as they are too noisy.
The procedure also struggles to predict the $f_\mathrm{esc}$ of noisy spectra (described in Section~\ref{sec:fesc_est}). Therefore 
we limit the our training and testing set to contain spectra with $m_\mathrm{AB,1500}<28.5$. We summarise the accuracy of our predicted $z$ in Table~\ref{tab:z_pred_summary}. The mean and standard deviation ($\sigma$) of the  relative error of all the testing set spectra is comparable to the values given in \citet{2016A&A...590A..31C}. We classify the spectra with $\left |\frac{\Delta z}{1+z}\right | \gtrsim 0.06$~\SG{$(\approx 3\sigma_{\frac{\Delta z}{1+z}})$} as outliers. In general, we find very few such outlier galaxies (less than \SG{1} per cent).

As the unshifting process is dependent on the SNR of the spectra, we consider spectra with similar noise for $t_\mathrm{exp}=1.5$ h. From figure~2 of \citet{Jensen16}, we infer that the spectra with $m_\mathrm{AB,1500}<27$ will have similar SNR to that of the spectra with $m_\mathrm{AB,1500}<28.5$ and $t_\mathrm{exp}=24$ h. As  shown in the right-hand panel of Figure~\ref{fig:z_est}, the $t_\mathrm{exp}=1.5$ h case also results in similar redshift scatter. 
The boxiness in the scatter plots of Figure~\ref{fig:z_est} is due to the process of making the wavelength sampling uniform during the unshifting of all the spectra. Since there is significant overlap of data points close to the $z^\mathrm{true}$ = $z^\mathrm{pred}$ line, these plot highlights the extreme outliers in the redshift determination but does not allow their relative occurrence to be inferred. 

\subsection{Escape fraction estimation}
\label{sec:fesc_est}

\begin{figure*}
  \centering
  \includegraphics[width=0.95\textwidth]{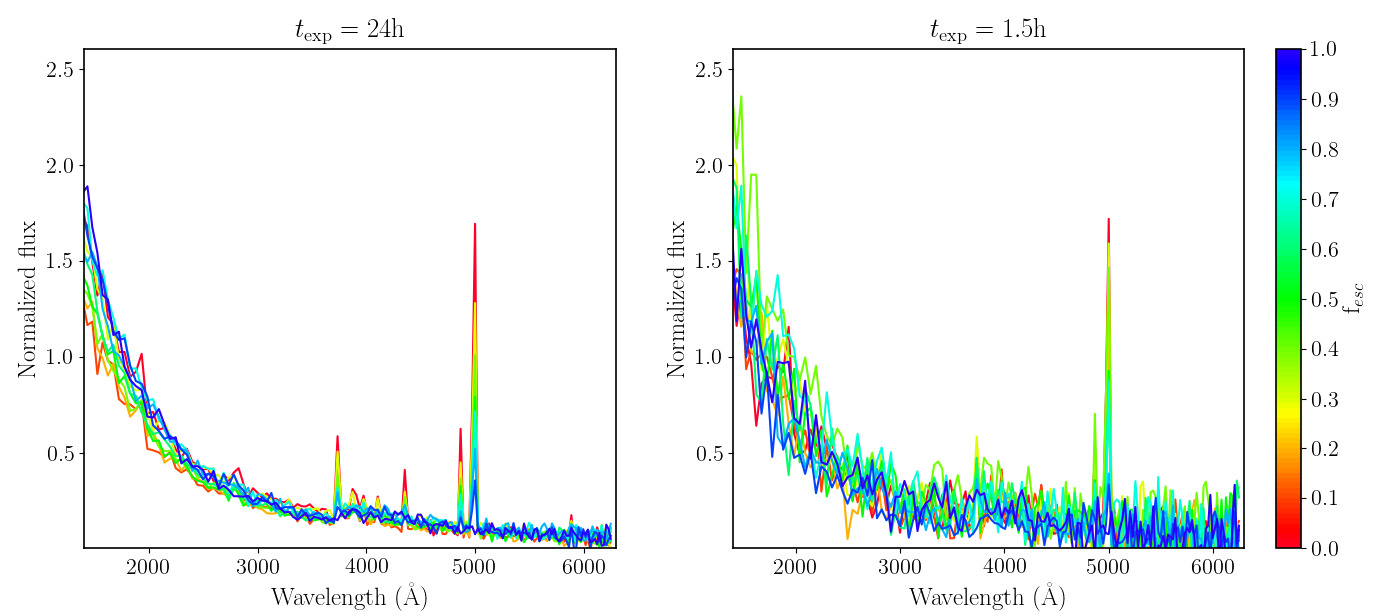}
  \caption{Spectra from galaxies with \SG{$m_\mathrm{AB,1500}\approx 27$ and} different escape fraction ($f_\mathrm{esc}$) after unshifting from $z\approx 7$. The left panel shows the spectra with 24 h exposure time while the right panel gives the ones with 1.5 h exposure time. Certain strong nebular lines, such as the H$\beta$ and [OIII] show a significant correlation with $f_\mathrm{esc}$. As expected, spectra observed with higher exposure time (24 h) allow for more clear-cut detection of emission lines.}
  \label{fig:spectra_bothexp}
\end{figure*}

\begin{figure*}
  \centering
  \includegraphics[width=0.85\textwidth]{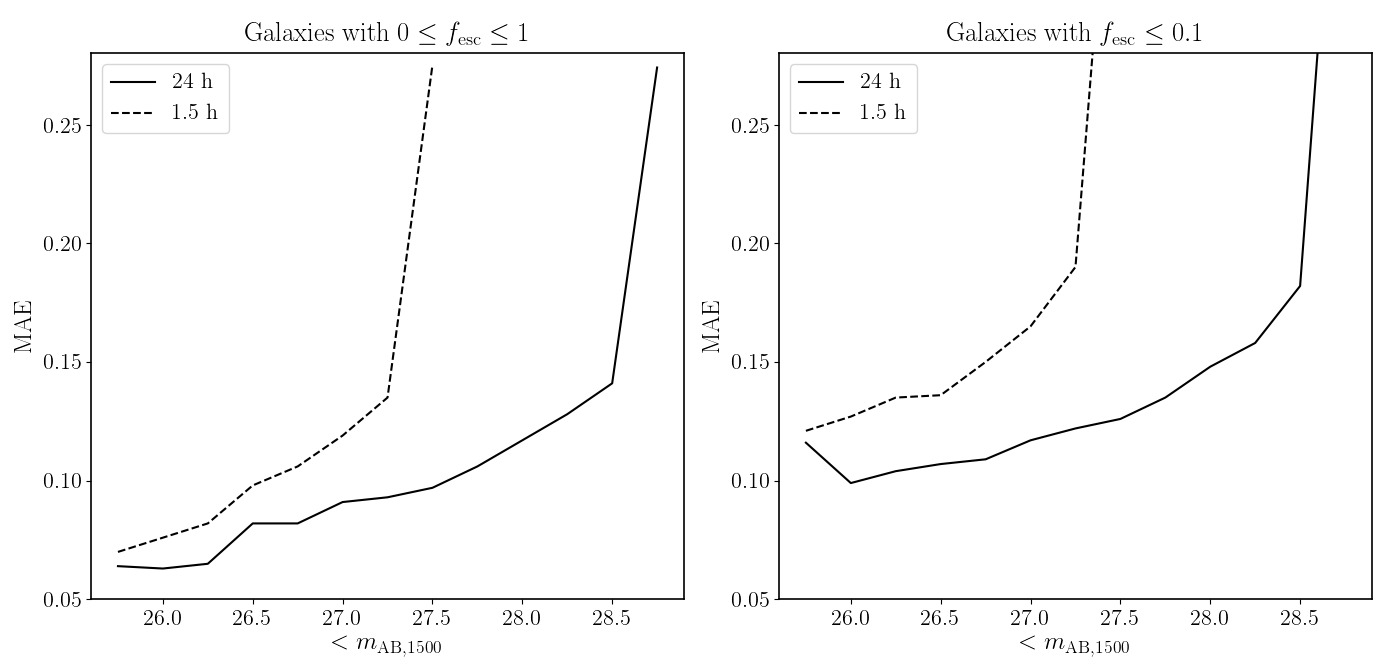}
  \caption{Mean absolute errors (MAEs) on $f_\mathrm{esc}$ calculated for the testing set with different upper limits on $m_\mathrm{AB,1500}$. The dashed and solid line shows the MAE for the data sets simulated with 1.5 h and 24 h exposure times, respectively. \SG{In the left panel, we show the estimate the MAEs for the testing set containing galaxies with a uniform $f_\mathrm{esc}$ distribution in the range $0\leq f_\mathrm{esc} \leq 1$}.  We find that our framework works well for the galaxies with $m_\mathrm{AB,1500} < 27$ mag for both exposure times. With $t_\mathrm{exp}=24$ h, the method can be extended to include even fainter galaxies ($m_\mathrm{AB,1500} < 28.5$ mag). \SG{In the right panel, we show the estimated MAEs for the testing set containing galaxies with $f_\mathrm{esc} \leq 0.1$. The value for MAEs for testing sets with various upper limits on $m_\mathrm{AB,1500}$ are larger compared to the ones shown in the left panel. The estimated MAEs are larger than 0.1 for all the cases. Thus, our procedure cannot clearly separate the galaxies with low escape fractions.
  }}
  \label{fig:mae_mcut}
\end{figure*}

\begin{figure*}
  \centering
    \includegraphics[width=0.85\textwidth]{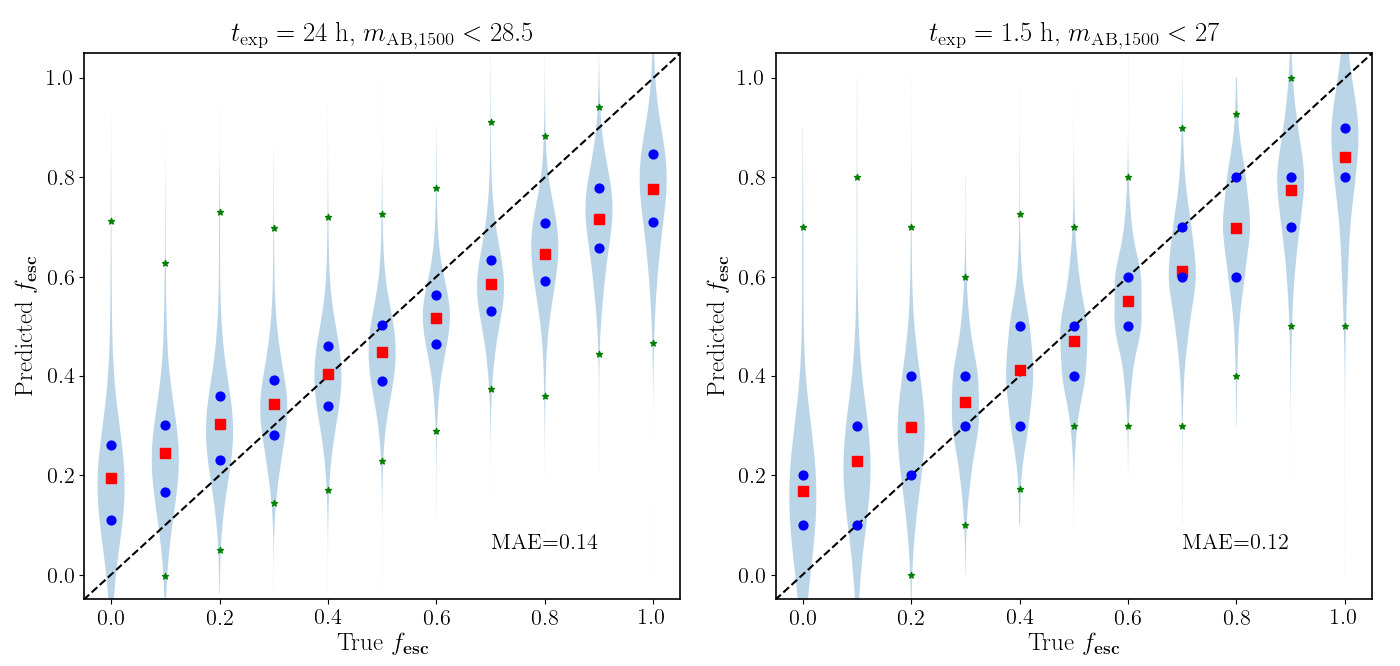}
    \caption{Prediction of $f_\mathrm{esc}$ for galaxies with $z=6$--10 as illustrated by violin plots, where the width of the violin indicates the probability density. The vertical violins indicate the probability with which each true $f_\mathrm{esc}$ value will be predicted. The panels on the left and right correspond to 24 h and 1.5 h exposure times, respectively. The black-dashed line represents $f_\mathrm{esc}^\mathrm{true}$ = $f_\mathrm{esc}^\mathrm{pred}$. The red squares indicate the median, the two blue circles the 25 and 75 percentiles and the two green stars the 1 and 99 percentiles of each distribution.  The black dashed line roughly passes through the region between the 25 and 75 percentiles for both exposure times. The deviation of the red data points from the black dashed lines shows that predictions $f_\mathrm{esc}$ are biased somewhat high for the lower $f_\mathrm{esc}^\mathrm{true}$ and low for the higher $f_\mathrm{esc}^\mathrm{true}$.
    }
  \label{fig:fesc_est}
\end{figure*}

After we have unshifted the observed spectra to the rest frame wavelength, the model fitted by LASSO to predict the escape fraction $f_\mathrm{esc}$ is similar to \citet{Jensen16}. In Figure~\ref{fig:spectra_bothexp}, we show the rest-frame spectra for both exposure times after unshifting from $z\approx 7$. A high $f_\mathrm{esc}$ is clearly correlated with weak emission lines. The $f_\mathrm{esc}$ parameter depends on the strength of the nebular emission lines because the enveloping gas cloud causing the lower $f_\mathrm{esc}$ absorbs photons and re-emits through recombination lines. In these simulations, a clear correlation between the UV slope and $f_\mathrm{esc}$ also emerges, but this may be overthrown for dust attenuation recipes other than that adopted here \citep{Zackrisson17}.

Although the rest-frame spectra for the shorter exposure time appear more noisy (right hand panel of Figure~\ref{fig:spectra_bothexp}), we can still clearly see the correlation between the $f_\mathrm{esc}$ and some prominent emission lines. A weak correlation with the UV slope is also visible. However, some of the weak emission lines are hidden in the noise. This will complicate the prediction of the $f_\mathrm{esc}$ as LASSO will either see noise peaks as emission lines or will not be able to identify some emission lines. When we compare the coefficients fitted by LASSO for both data sets, we see that there are fewer non-zero coefficients for shorter exposure times. In this case some emission lines are missed by LASSO, which results in worse estimates for $f_\mathrm{esc}$. 
The above results are for spectra with $z\approx 7$. The correlations are stronger for spectra at $z< 7$, but at higher $z>7$, the correlation between the measured UV slope and $f_\mathrm{esc}$ is so much worse for the $t_\mathrm{exp}$ = 1.5 h case that it will significantly affect the accuracy of the $f_\mathrm{esc}$ prediction.

LASSO fits a model for $f_\mathrm{esc}$ that is similar to the ones shown in \citet{Jensen16}. The nebular emission lines have a strong correlation with the $f_\mathrm{esc}$. The coefficient for the \textit{shift} value is zero as  $f_\mathrm{esc}$ is independent of this parameter. Our data set consists of a mix of bright and faint galaxies and our method struggles when there are more faint galaxies in the training and testing set. Therefore we limit the magnitude of the spectra when we analyse them. The limit imposed on the data set is the same for both training and testing sets. In order to decide on the upper limit, we plot the mean absolute error (MAE) of our $f_\mathrm{esc}$ prediction against various maximum $m_\mathrm{AB,1500}$ of the training and testing data set in \SG{the left panel of} Figure~\ref{fig:mae_mcut}. 

The $f_\mathrm{esc}$ prediction pipeline works well for galaxies with $m_\mathrm{AB,1500}<27$ for both $t_\mathrm{exp}$. We can analyse spectra from even fainter galaxies with larger $t_\mathrm{exp}$. Throughout the rest of this paper, we will use $m_\mathrm{AB,1500}<28.5$ and $<27.0$ as the faint limits for for $t_\mathrm{exp}$ of 24 h and 1.5 h respectively.

We show the prediction of $f_\mathrm{esc}$ from the spectra in our testing set in Figure~\ref{fig:fesc_est} using violin plots. The width of the violins represent the probability of prediction of a certain value. We determine this width using a non-parametric density estimator based on the Parzen--Rosenblatt window method \citep{rosenblatt1956remarks,parzen1962estimation}. The vertical violins in both the panels  give the probability distribution function (PDF) of the predicted values for each true $f_\mathrm{esc}$. The MAE is in agreement with \citet{Jensen16} and the mean values of the predicted $f_\mathrm{esc}$ also reproduce the trend shown in \citet{Jensen16}, despite the fact that the algorithm now also needs to estimate the redshift before attempting an $f_\mathrm{esc}$ prediction.
. 

The vertical violins for the smaller exposure time are longer, which can be inferred from more distant 1 and 99 percentile markers. However, the MAE for $f_\mathrm{esc}$ prediction on this testing set is 0.11.
The 25 and 75 percentile markers are seen at similar values. This suggests that even though the spectra with small exposure times will result in some spurious outliers, most predictions will be close to the true values. The black dashed lines represents $f^\mathrm{true}_\mathrm{esc}$ = $f^\mathrm{pred}_\mathrm{esc}$ in all the violin plots. We see that the black dashed line falls within the 25 and 75 percentile markers for most $f_\mathrm{esc}$ values in both data sets. The violins at the extreme values of $f^\mathrm{true}_\mathrm{esc}$ show less accurate predictions. The black dashed line passes through the 1 and 25 percentile marker for low $f^\mathrm{true}_\mathrm{esc}$, but the 75 and 99 percentile marker for high $f^\mathrm{true}_\mathrm{esc}$. We observe this effect more in the data set that contains fainter sources. In case of the spectra with high $f_\mathrm{esc}$, the noise spikes are sometimes confused as spectral feature by LASSO resulting in prediction of lower $f_\mathrm{esc}$. Therefore the mean value of the of $f^\mathrm{pred}_\mathrm{esc}$ is always lower at the high end of $f_\mathrm{esc}$. In case of the spectra with low $f_\mathrm{esc}$, the noise hides the spectral features, thus, giving higher $f^\mathrm{pred}_\mathrm{esc}$.

\subsection{Data sets with few high-$f_\mathrm{esc}$ galaxies}

\begin{figure*}
  \centering
  \includegraphics[width=0.90\textwidth]{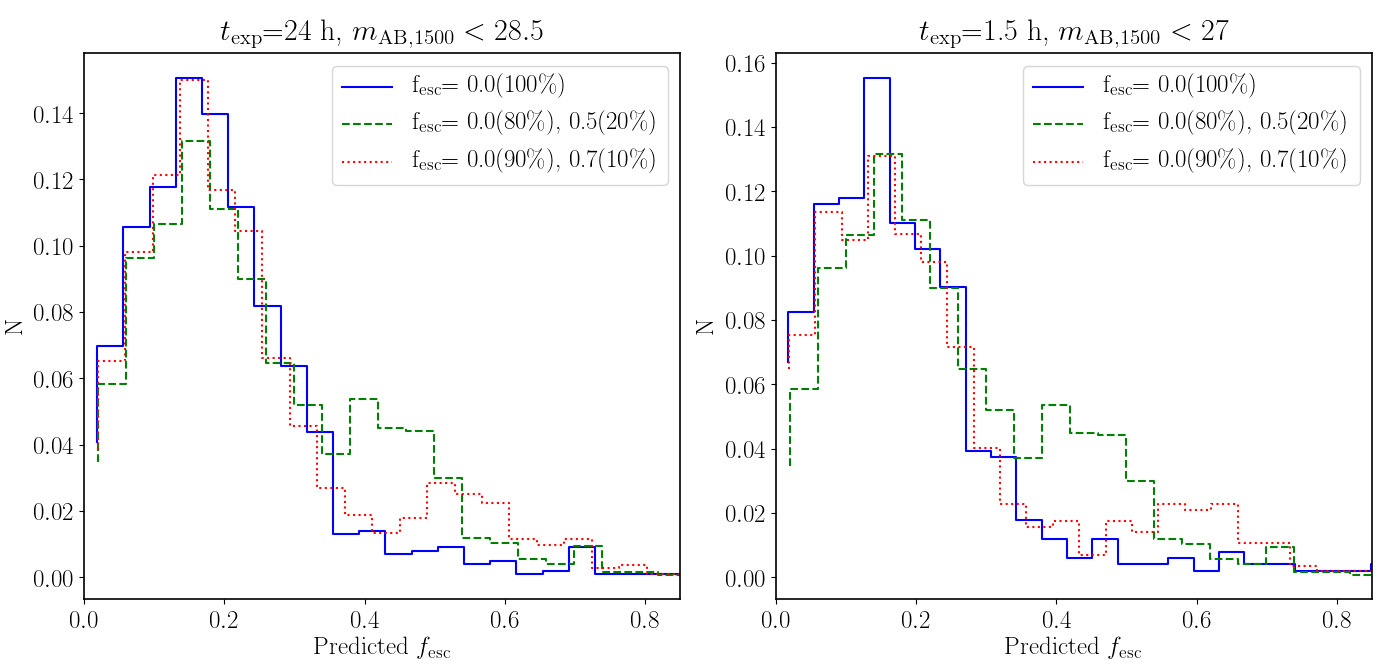}
  \caption{Histograms of predicted $f_\mathrm{esc}$ for three scenarios in which the $z=6$--10 galaxy population is assumed to contain just a minor fraction of high-$f_\mathrm{esc}$ objects (as described in the legend). The left and right panels show the predictions for $t_\mathrm{exp}$ = 24 h and 1.5 h respectively. When the data set contains only galaxies with $f_\mathrm{esc}=0$, the histogram displays a single peak. However, the presence of a subpopulation of high-$f_\mathrm{esc}$ cases makes the histogram bimodal with a second peak close to the $f_\mathrm{esc}$ of this high-$f\mathrm{esc}$ population.
  }
  \label{fig:fesc_est_caseE}
\end{figure*}

The testing set we used for illustrating the performance of our framework in Section~\ref{sec:fesc_est} has a uniform mix of galaxy spectra with various $f_\mathrm{esc}$. However, a more realistic data catalogue of JWST observations may contain very few high-$f_\mathrm{esc}$ ($f_\mathrm{esc}\geq 0.5$) cases. To test the performance of our procedure in this situation, we consider two $f_\mathrm{esc}$ distribution scenarios and compare these with the case where all galaxies have $f_\mathrm{esc}=0$ (which can be considered a proxy for the case where all galaxies have $f_\mathrm{esc}<0.1$). In the first scenario, 80 per cent of the galaxies have $f_\mathrm{esc}=0$ and 20 per cent have $f_\mathrm{esc}=0.5$. In the second scenario, $90$ per cent of the galaxies have $f_\mathrm{esc}=0$ and 10 per cent have $f_\mathrm{esc}=0.7$.

In Figure~\ref{fig:fesc_est_caseE}, we show the histograms of the predicted $f_\mathrm{esc}$ values for the these cases when $t_\mathrm{exp}$ = 24 h (left panel) and $t_\mathrm{exp}$ = 1.5 h (right panel). The $f_\mathrm{esc}$ histogram for both exposure times is unimodal when the data set contains just galaxies with a single $f_\mathrm{esc}$. However, when there is a mixture of different $f_\mathrm{esc}$, we see a second peak close to the $f_\mathrm{esc}$ value of the high-$f_\mathrm{esc}$ subpopulation. Hence, the procedure is able to identify the presence of a a minor fraction of  high-$f_\mathrm{esc}$ objects within the sample. As expected, the secondary, high-$f_\mathrm{esc}$ peak becomes more prominent when the fraction of high-$f_\mathrm{esc}$ objects is higher or when the $f_\mathrm{esc}$ value of the high-$f_\mathrm{esc}$ subpopulation is extremely high ($f_\mathrm{esc}=0.7$ in this example). 
The secondary peak is always skewed towards the lower $f_\mathrm{esc}$ values. The reason for the bias are the noise spikes, which was also described in the previous section.

\subsection{Data sets with only low-$f_\mathrm{esc}$ galaxies}

\SG{The procedure presented in this paper is mainly intended for identifying high-$f_\mathrm{esc}$ galaxies, and our $f_\mathrm{esc}$ predictions have relatively low accuracy in cases where $f_\mathrm{esc}$ is in the $f_\mathrm{esc}\approx 0$--0.1 range (Figure~\ref{fig:fesc_est}). To explore the limits of the method in this regime, we take a testing set containing galaxies with $f_\mathrm{esc} \leq 0.1$ and place various upper limits on  $m_\mathrm{AB,1500}$ to estimate the MAE of the predicted $f_\mathrm{esc}$. The right panel of Figure~\ref{fig:mae_mcut} shows the dependence of MAEs on the limiting $m_\mathrm{AB,1500}$ cut for both 24 h and 1.5 h exposure times. The resulting MAE values are clearly larger than in the case when all $f_\mathrm{esc}$ values are considered  (left panel of Figure~\ref{fig:mae_mcut}), and never smaller than $\Delta f_\mathrm{esc}\approx 0.1$. This implies that the $f_\mathrm{esc}=0$ and $f_\mathrm{esc}=0.1$ cases cannot be clearly separated. However, one may still be able to infer that a given observed object displays characteristics that are consistent with $f_\mathrm{esc}<0.2$--0.3.}


\section{Discussion}
\label{discussion}

\subsection{Modelling uncertainties}
The method described in this paper elucidates that JWST/NIRSpec observations have the potential for estimating the $f_\mathrm{esc}$ parameter and identifing high-$f_\mathrm{esc}$ candidates from $z\gtrsim 7$ galaxy samples. However, it must be stressed that this will only work in practice if the simulations on which the machine learning machinery is trained has been based on sufficiently realistic templates for actual EoR galaxies. \citet{Jensen16} showed that the procedure results in poor predictions of $f_\mathrm{esc}$ when the training and testing spectra are based on different simulations (a proxy for potential mismatches between simulations and real EoR galaxies). A high-priority task in the early years of JWST science will therefore be to test theoretical predictions concerning the ionizing flux of stellar population models, the dust distribution and the star formation history against the data. 

\subsection{Redshift uncertainties} 
Our procedure of unshifting mock JWST/NIRSpec spectra followed by LASSO gives a well constrained redshift prediction and this method is similar to the ones previously used for determining the redshifts of galaxies in large surveys such as the Baryon Oscillation Spectroscopic Survey (BOSS), the Sloan Digital Sky Survey (SDSS) and the 2-degree Field Lensing Survey. In these studies, redshifts are determined by cross-correlating the observations with a sample of observations with known redshifts \citep{2013MNRAS.433.2857M,2017MNRAS.465.4118J}. In contrast to their methods, however, we have a dictionary of simulated spectra in rest frame wavelengths comprising all possible $f_\mathrm{esc}$ scenarios (the GSS in our machinery). 

The caveat with this redshift determination method is that the GSS and the mock observations are created from the same simulation. The goal of the unshifting step is to have the spectral features in the same wavelength bin, and once real observations are at hand, we can follow the previous studies \citep{2013MNRAS.433.2857M,2017MNRAS.465.4118J} and hand-pick the best spectra (with prominent spectral lines) from the sample to unshift all the observed spectra.

\section{Summary}
\label{summary}
In this work, we have presented an $f_\mathrm{esc}$ prediction pipeline based on LASSO regression, the machine learning method introduced in \citet{Jensen16} to study EoR galaxy spectra. We apply our procedure to a set of simulated galaxy spectra with fluxes scaled to closely match the objects in lensed fields, which are already scheduled to be observed with JWST/NIRSpec. Our study shows that LASSO can both be used to \SG{determine the redshifts and constrain the escape fractions of high-$z~(\gtrsim 6)$ galaxies. For spectra observed with 24 h JWST/NIRSpec exposure at a resolution of $R=100$, we find that our machinery can predict the escape fractions of galaxies brighter than $m_\mathrm{AB,1500}\approx 27$ with a relative error $\Delta f_\mathrm{esc}\approx 0.09$. If we include fainter galaxies ($m_\mathrm{AB,1500}\lesssim 28.5$), the relative error $\Delta f_\mathrm{esc}$ becomes $\approx 0.14$. Our procedure can also be applied on noisier data set. In this study, we tested our procedure on simulated JWST/NIRSpec spectra observed with 1.5 h exposure at resolution of $R=100$. In this case, the escape fractions of bright galaxies ($m_\mathrm{AB,1500}\lesssim 27$) can be predicted with a relative error $\Delta f_\mathrm{esc}\approx 0.12$.}

\SG{Our method is primarily intended for identifying EoR galaxies with very high levels of LyC leakage, and our study shows that LASSO can both be used to simultaneously identify redshifts and single out high-$f_\mathrm{esc}$ candidates ($f_\mathrm{esc}>0.5$) from a JWST/NIRSpec data set.} In an idealised case where the training set simulations provide a perfect match to real EoR galaxies, the algorithm would -- in a sample of galaxies with the same redshift and brightness distribution as the current photometric candidates in the A2744 and M0416 cluster fields -- identify $\lesssim 5$ per cent false-positive $f_\mathrm{esc}\gtrsim 0.5$ candidates in a scenario where all galaxies in fact have $f_\mathrm{esc}< 0.1$. 
Similarly, the algorithm would be able to correctly identify $\approx 80$ per cent of the high-$f_\mathrm{esc}(\gtrsim 0.5)$ candidates even when the fraction of these high leakers is as low as 20 per cent in the whole sample. 
When the sample contains a subset of high-$f_\mathrm{esc}$ candidates, the distribution of the predicted $f_\mathrm{esc}$ will be bimodal, which will a clear evidence of presence of high leakers in the sample.

In coming years, the main issue is of course whether contemporary simulations are able to provide sufficiently realistic representations of actual EoR galaxies to be useful as training sets in this respect. A high-priority task in the early days of JWST science will therefore be to test simulations against NIRSpec data in search for potential discrepancies of a type that may bias the results \citep[for suggested diagnostic tests of this type, see][]{Zackrisson17,Binggeli18}.

\section*{Acknowledgements}
We acknowledge Matthew Hayes, Garrelt Mellema and Hannah Ross for useful discussions. We also thank Tommaso Treu for helpful information about Through the looking GLASS. 
EZ acknowledges funding from the Swedish National Space Board.





\bibliographystyle{mnras}
\bibliography{refs}




\bsp	
\label{lastpage}
\end{document}